\documentclass[aps,prl,twocolumn]{revtex4-1}
\usepackage{graphicx}
\usepackage{dcolumn}
\usepackage{bm}
\usepackage{multirow}
\usepackage{ulem}
\usepackage{color}

\begin{document}

\title{Optical-lattice based Cs active clock with continual superradiant lasing signal}

\author{$^1$Duo Pan}
\author{$^2$Bindiya Arora}
\author{$^3$Yan-mei Yu}
\email{ymyu@aphy.iphy.ac.cn}
\author{$^4$B. K. Sahoo}
\email{bijaya@prl.res.in}
\author{$^1$Jingbiao Chen}
\email{jbchen@pku.edu.cn}

\affiliation{$^1$State Key Laboratory of Advanced Optical Communication Systems and Networks, Department of Electronics, Peking University, Beijing 100871, China}
\affiliation{$^2$Department of Physics, Guru Nanak Dev University, Amritsar, Punjab 143005, India}
\affiliation{$^3$Beijing National Laboratory for Condensed Matter Physics, Institute of Physics, Chinese Academy of Sciences, Beijing 100190, China}
\affiliation{$^4$Atomic, Molecular and Optical Physics Division, Physical Research Laboratory, Navrangpura, Ahmedabad 380009, India}

\begin{abstract}
We demonstrate state-of-the-art technique of an active clock to provide a continuous superradiant lasing signal using an ensemble of trapped Cs
atoms in the optical lattice. A magic wavelength of the proposed $|7S_{1/2}; F=4, M_F=0\rangle \rightarrow |6P_{3/2}; F=3, M_{F}=0 \rangle$ clock
transition in Cs atom is identified at 1181 nm for constructing the optical lattice. Pertinent optical lines are also found for pumping and
repumping atoms from their ground states. A fractional uncertainty about 10$^{-15}$ level to the clock frequency has been predicted by carrying out
rigorous calculations of several atomic properties. The bad-cavity operational mode of the active clock is anticipated to improve its
short-term stability remarkably by suppressing intrinsic thermal fluctuations. Thus, a composite clock system with better in both short-term
and long-term stabilities can be built by combining the above proposed active clock with another high-accuracy passive optical clock.
\end{abstract}
\date{\today}

\maketitle

 It is well known that owing to its natural immunity against the cavity¡¯s thermal noise floor, hydrogen masers offer better instability among the microwave atomic clocks within the first 30 seconds of frequency measurement, referred to as its short-term stability. On the other hand, the short-term stabilities of the existing high-precision optical passive clocks have strict requirement for are constrained by the thermal noise of the local oscillators' super-cavities instabilities of the local oscillators and the thermal noise floor of reference cavities. With the development of cryogenic optical cavities thermal noise has been continuously reduced \cite{Matei2017,Kessler2011,Zhang2017}, and the short-term instability of local oscillators as low as 4E-17@1s has been achieved. This has brought significant improvement in the short-term stability of passive optical clocks \cite{Yb2015,Sr2018Japan,Al+2017,Yb+2016,Tm2019,Michal2018}. In the meantime, a new direction has emanated to improve short-term stability for an optical clock. By manifesting the working principle of hydrogen maser, building an active optical clock is steadily gaining the ground \cite{Meiser-PRL-2009,Kazakov-PRA-2017,KuppenLinewidth,Norcia-PRX-2018,Laske-PRL-2019,YuDS-PRL-2007}. The underlying principle of an active optical clock
lies in its robustness to infer the clock frequency directly from the superradiant signal of an ensemble of atoms placed inside a ``bad-cavity''. In principle, the active optical clock can achieve short-term instabilities well below the cavities¡¯ thermal noise limit and at the same time it can relax the requirement for the cryogenic ultra-stable cavity, thus reducing the systematic complexity.
The short-term stabilities of these clocks can be improved appreciably by suppressing photon shot-noises through continuous pumping and output lasing signals. The
previously proposed two-level active optical clocks \cite{Norcia-PRX-2018,Laske-PRL-2019} are compelled to produce signals intermittently to overcome
interference between the applied pumping pulse with the output clock lasing signal. The stability and accuracy of a continuously signal producing
proposed in three-level clock \cite{Meiser-PRL-2009} are critically obstructed by the light shift caused by the pumping laser. To explore in this direction further, we had analyzed four atomic levels accessible by optical lasers in an ensemble of thermal Cs atoms for a possible continuously signal producing active optical clock in Ref. \cite{Pan-02}. However, this is also subjected to large systematics due to light shifts and collisional decoherence.

\begin{figure*}[t]
\includegraphics[width=5.5cm,height=4cm]{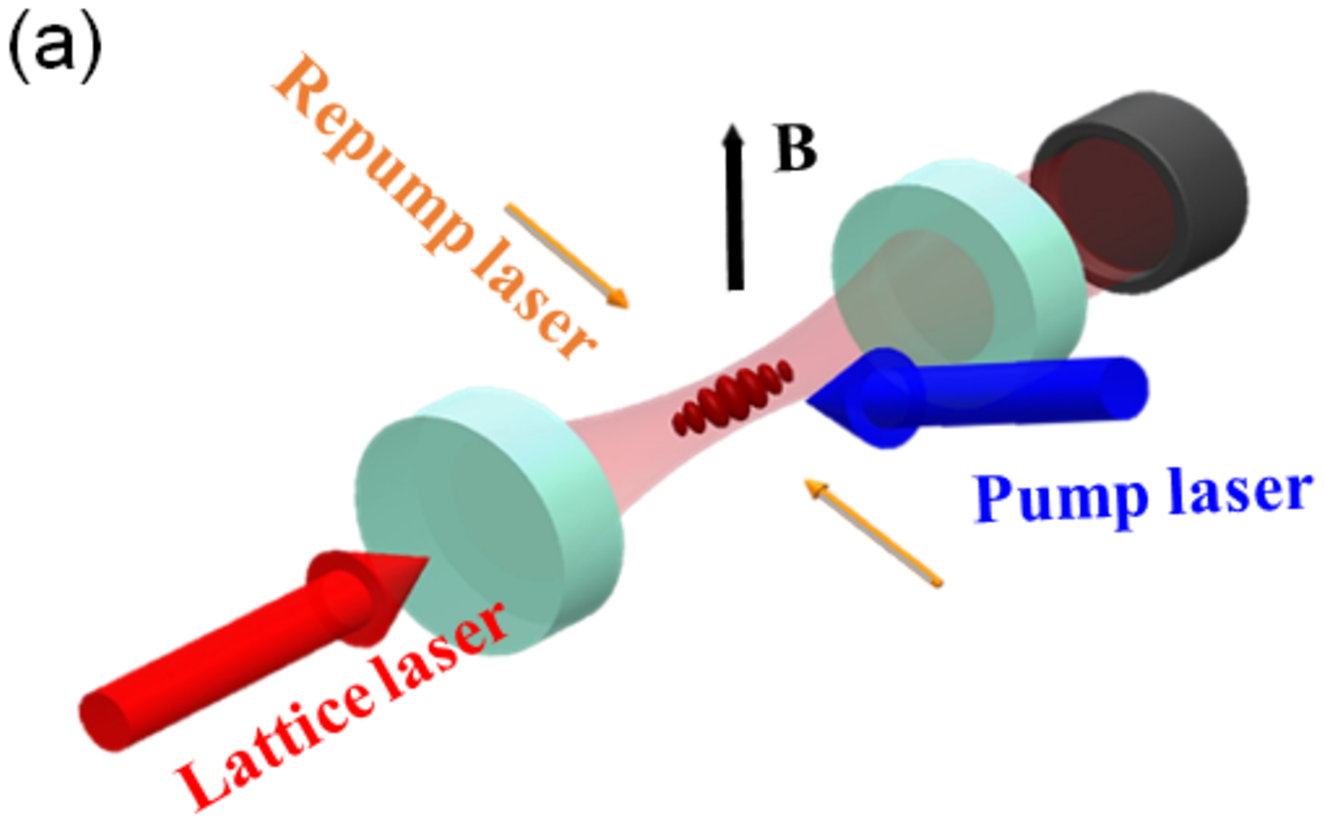}
\includegraphics[width=5.5cm,height=4cm]{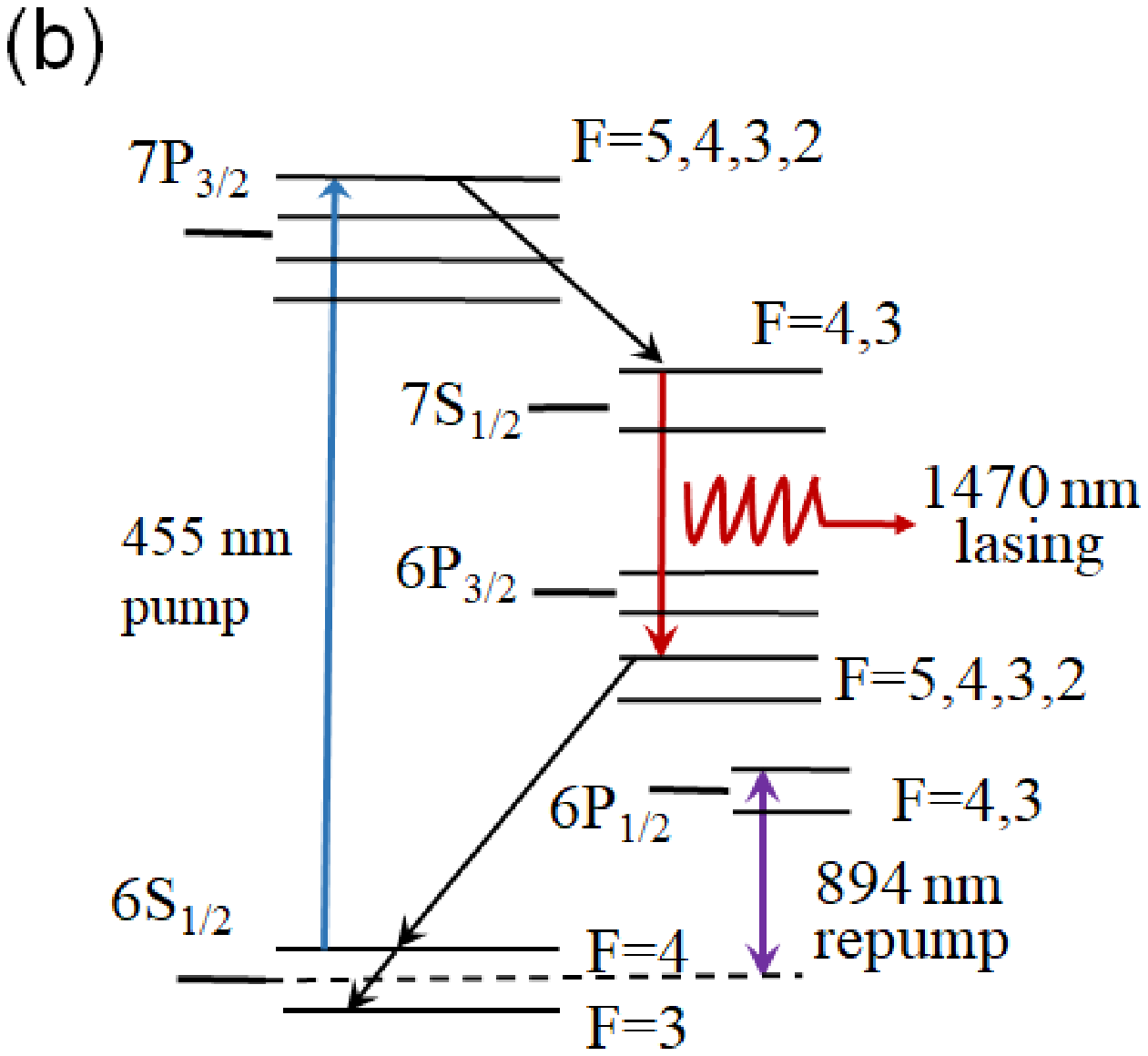}
\includegraphics[width=5.5cm,height=4cm]{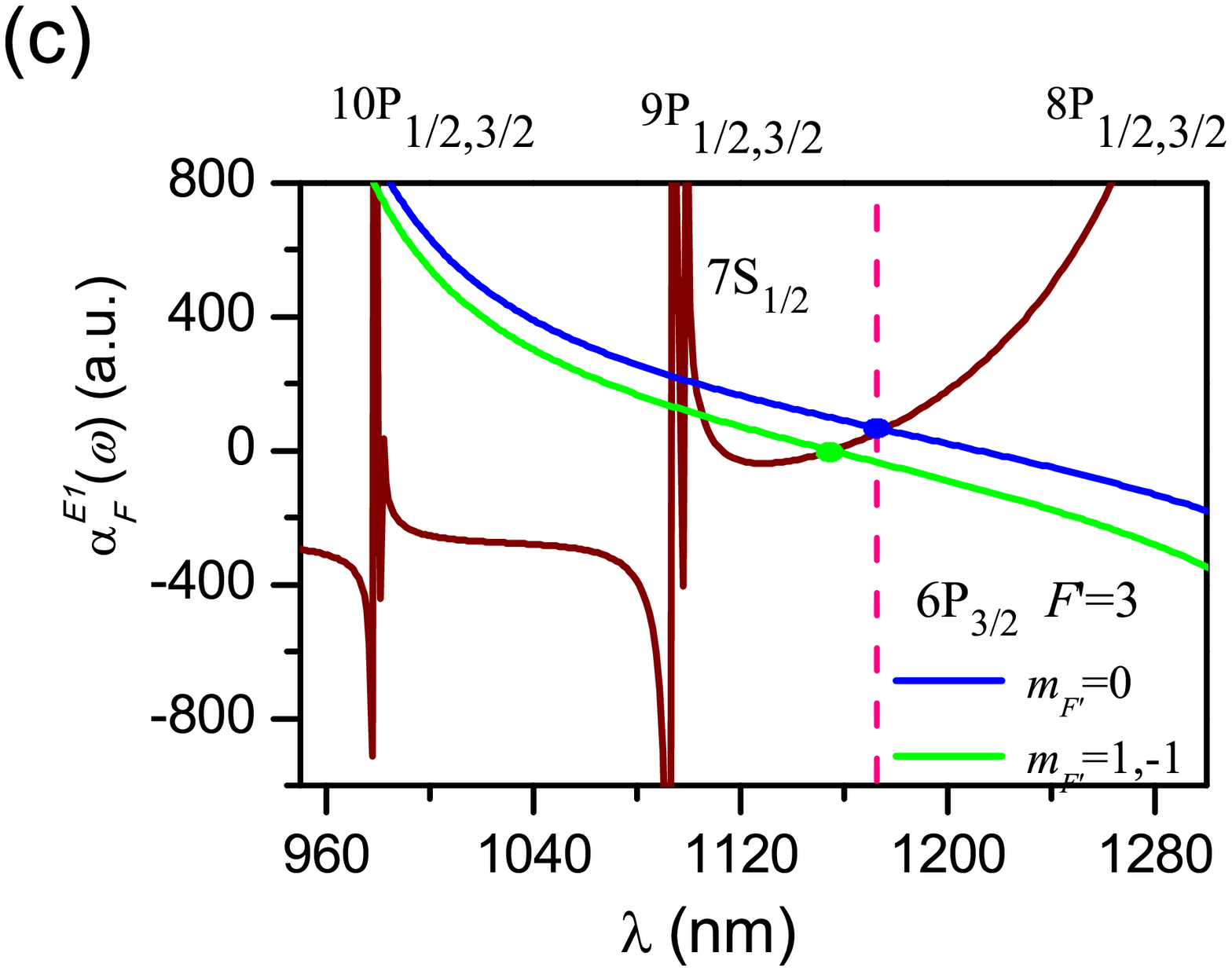}
\caption{\label{lasing}
Schematic depictions of (a) experimental set-up, in which cold Cs atoms are held in an 1D optical lattice situated inside a high finesse cavity near
magic wavelength ($\lambda_m$) of $|7S_{1/2}; F=4, M_F=0 \rangle \rightarrow |6P_{3/2}; F=3, M_F=0 \rangle$ clock transition of the proposed
four-level active clock, (b) the relevant energy levels for clock, pumping and repumping transitions, and (c) plot for $\alpha^{E1}_F(\omega)$ values
(in a.u.) of the clock levels against wavelength $\lambda$ (in nm) to infer required $\lambda_m$. Resonance lines for the $7S_{1/2}$ state are marked
on the top of plot.}
\end{figure*}

In this Letter, we expound feasibility of an active optical clock using four suitable hyperfine levels of trapped cold Cs atoms, instead of atomic
energy levels of thermal Cs atoms proposed in Ref. \cite{Pan-02}, in a red-detuned one-dimensional (1D) optical lattice at a magic wavelength
($\lambda_m)$ of the $|7S_{1/2}; F=4, M_F=0\rangle \rightarrow |6P_{3/2}; F=3,M_{F}=0 \rangle$ clock transition. The schematic layout of our
experimental set-up and the associated hyperfine levels are shown in Figs. \ref{lasing}(a) and (b), respectively. The unique advantages
of adopting this technique with appropriate choices of hyperfine levels are at least three folds over the previous active optical clocks, such as it
moderates interference between the pumping laser and the superradiant lasing signal, and minimizes light shifts and collisional decoherence to the
hyperfine levels and provides sufficient time to conduct the experiment meticulously due to strong confinements of atoms in the optical lattice.
A sufficiently weak magnetic field, of about ${\cal B}=10^{-7}$ T, is applied to break the $M_F$ degeneracy of the clock transition and a
$\pi-$polarized pumping laser is locked at 455 nm for populating Cs atoms to the $|7P_{3/2}; F=5,M_F=0\rangle$ level from the $|6S_{1/2}; F=4,
M_F=0\rangle$ level of the ground state. Subsequently, these atoms decay to any of the $M_F=-1,0$, or $1$ sublevels of the $F=4$ hyperfine level of
the $7S_{1/2}$ state through spontaneous emission. At the same time, a mode of bad-cavity is adjusted to couple with the clock transition, which is
further locked to a super-cavity \cite{Pan-02}. The objective of this step is to attain self-gaining stimulated emission from the above transition
in the bad-cavity mode and produce a superradiant lasing. A polarization selector inside the cavity (not shown in the figure) helps to establish a
superradiant lasing signal in the above transition through the allowed $(M_{F}\to M_{F}) \ \equiv \ (0\to0)$, $(1\to1)$, and $(-1\to-1)$ decay
channels. Then, atoms come down to either of the $F=3$ or $F=4$ hyperfine levels of the ground state. A 894 nm laser is applied at this stage to
repump the atoms from all other magnetic sub-levels to the desired $|6S_{1/2}; F=4, M_F=0\rangle$ level of the ground state by using a dual-laser
\cite{dualrepump}. The combined pumping action, spontaneous emission, superradiant emission, and spontaneous decay channels follow the path of the
$6S_{1/2} \rightarrow 7P_{3/2} \rightarrow 7S_{1/2} \rightarrow 6P_{3/2} \rightarrow 6S_{1/2}$ transitions successively to populate the $|6S_{1/2};
F=4, M_F=0 \rangle$ level in the end of a complete cycle to perpetuate a continuous loop of clock operation. There is a probability that a fraction
of atoms can populate in the $5D_{3/2,5/2}$ states through spontaneous emission from the $7P_{3/2}$ state, but they will decay to the $6P_{1/2,3/2}$
states within 0.1 $\mu$s. Thus, the superradiant lasing is not going to be affected by such process. Since interferences between the (re)pumping
laser and the clock lasing signal do not occur in this procedure, a continuous signal of superradiant lasing at 1470 nm can be achieved.

\begin{table*}[t]
\caption{Properties of Cs atoms and optical trap at 1181 nm magic wavelength. Here, $\tau$, $\alpha_J^{E1}$ and $\rho$ represent natural lifetimes, atomic dipole polarizabilities, and populations, respectively, of investigated states in Cs atom, while $U_{0}$, $\Gamma_{sc}$ and $\tau_{L}$
correspond to trapping potential, scattering rate and lifetimes of atomic states due to the scattered photons, respectively, in the 1D optical
lattice for the operational laser intensity $I_{op} = 20$ kW/cm$^2$. Uncertainties to the $\alpha_J^{E1}$
values are quoted explicitly, which are used for estimating the clock frequency shift. \label{tab:magic}}
{\setlength{\tabcolsep}{18pt}
\begin{tabular}{ccccccc }\hline\hline
State     & $\tau$($\mu$s) & $\alpha_J^{E1}$(a.u.)& $U_{0}$(kHz) &   $\rho$(\%)    &$\Gamma_{sc}$(kHz/s) & $\tau_L$ (s)\\\hline
$6S_{1/2}$& $\infty$       &  846(11)             &$-2957$       &    27.1         &2.6                  &1137          \\[+1ex]
$5D_{5/2}$& 1.37           &  793(38)             &$-2772$       &    33.9         &21.5                 &129           \\[+1ex]
$5D_{3/2}$& 0.97           & 730(37)              &$-2551$       &    3.1          & 13.0                &196           \\[+1ex]
$7P_{3/2}$& 0.11           & 8045(4388)           &$-28119$      &    24.1         &0.1                  &$>10^5$        \\[+1ex]
$7S_{1/2}$& 0.05           & 73(10)               &$-255$        &    5.5          &0.3                  &850           \\[+1ex]
$6P_{3/2}$& 0.03           & -381(32)             &$-255$        &    2.7          &6.8                  &38            \\\hline\hline
\end{tabular}}
\end{table*}

An essential feature of our proposed experimental technique is to trap Cs atoms at a $\lambda_m$ of the clock transition in an optical lattice. The most convenient choice is to use a $\pi$-polarized laser for creating trapping potential, which can cause the ac-Stark shift in a hyperfine $| F, M_F \rangle$ level
of atomic state as $\Delta E_F(\omega)= - \frac{1}{4}\alpha^{E1}_F(\omega){\cal E}^2$ with amplitude of electric field ${\cal E}$ of trapping
laser and dynamic electric dipole (E1) polarizability of hyperfine level $\alpha^{E1}_F(\omega)$ for laser frequency $\omega$. The $\alpha^{E1}_F(\omega)$ values of the hyperfine levels of the $7S_{1/2}$ and $6P_{3/2}$ clock states are determined by calculating their atomic-state polarizability $\alpha^{E1}_J(\omega)$ values using the same method as in Refs. \cite{cspol1,cspol2}. For this purpose, we have used the experimental energies and E1 matrix elements either from the literature \cite{Toh-PRL-2019,Damitz-PRA-2019,Toh-PRA-2019} or by calculating them using a relativistic all-order method \cite{rcc}. Uncertainties to the
$\alpha^{E1}_{F}$ values are determined from the accuracies of the E1 matrix elements. The $\lambda_m$ values of a transition corresponds to its
null differential $\alpha^{E1}_F(\omega)$ values. We plot the $\alpha^{E1}_F(\omega)$ values of the hyperfine levels of the $7S_{1/2}$ and $6P_{3/2}$ states
in Fig. \ref{lasing}(c) for the optical range of 950 $-$ 1300 nm to locate possible $\lambda_m$ values. The intersections of the polarizability
curves mark the presence of a suitable $\lambda_m$ at 1181 nm. The inferred $\alpha^{E1}_J(\omega)$ values of different states at $\lambda_m=1181$ nm are quoted in atomic units (a.u.) in Table \ref{tab:magic}. At 1181 nm, $\alpha^{E1}_{F}$ is 73(10) a.u. for the $7S_{1/2}, F=4$ upper clock state, and $\alpha^{E1}_F=$73(36) a.u. for the $6P_{3/2}, F=3$ lower clock state, as determined by its scalar and tensor $\alpha^{E1}_J$ values of -381(32) a.u. and 629(13) a.u., respectively.

The 455 nm pumping laser can populate Cs atoms from the ground state on the excited state, simultaneously, the 894 nm repumping laser is used
for Sisyphus cooling. The temperature of the trapped cold Cs atoms is determined by balancing the heating rate and the cooling effect;
$k_BT_c=(D_p+D_p')/\alpha_{cool}$, where $D_p$ and $D_p'$ are the momentum diffusion coefficients of the pumping and repumping lasers, respectively,
and $\alpha_{cool}$ is the cooling coefficient due to the 894 nm laser. It yields $D_p=\frac{1}{2} \hbar^2k^2\Gamma_{455}=1.9 \times 10^{-48}
\rm{kg^2m^2}/\rm{s}^3$ \cite{Doppler} with the Rabi frequency $\gg \Gamma_{455}$, wherein $k$ and $\Gamma_{455}$ are the wave vector and spontaneous
decay rate, respectively, of the pumping transition and $\hbar$ is the Planck constant. We estimate $D_p'=9.1 \times 10^{-48} \rm{kg^2 m^2}/\rm{s}^3$
for a repumping laser intensity of 2 mW/cm$^2$ and detuning $\delta=30$ MHz using the formula $D_p'=(7/10)\hbar^2{k'}^2\Gamma_{894} s_0+(3/4)\hbar^2{k'}^2
(\delta^2/\Gamma_{894})s_0$ \cite{Dalibard1989Laser,Dalibard1984Potentialities}, where $s_0= \frac{\Omega^2/2} {\delta^2+\Gamma_{894}^2/4}$ is saturation parameter, wave vector $k'$, Rabi frequency $\Omega$ and spontaneous decay rate $\Gamma_{894}$ of the repumping transition. Similarly,
$\alpha_{cool}= 29 \hbar^2 {k'}^2 \Gamma_{894}/(50 k_BT_0)=2.2 \times 10^{-20}$ kg/s is obtained using the reported Sysphus cooling temperature
$T_0=30 \ \mu K$ \cite{Hou1998Experimental}. Accounting for the above parameters, it yields the cooling temperature of the system $T_c=35 \ \mu$K, which is low enough to ensure effective loading of Cs atoms into the lattice.

\begin{figure}[t]
\begin{center}
\includegraphics[width=7.5cm,height=5cm]{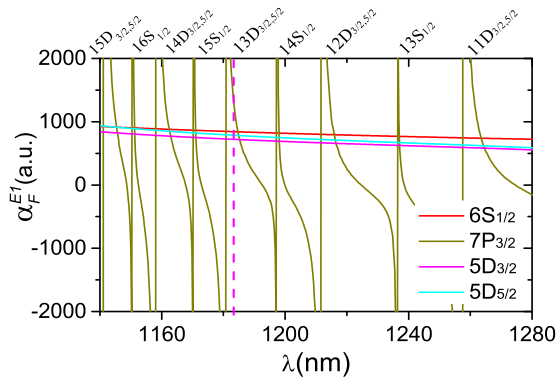}
\end{center}
\vspace{-6mm}
\caption{Plot for $\alpha^{E1}_F(\omega)$ values (in a.u.) of hyperfine levels used for pumping and repumping transitions against
wavelength (in nm). Resonance lines of the $7P_{3/2}$ state are labeled on the top of plot. The dashed line shows the position of the
$\lambda_m$ value (i.e. 1181 nm) for the optical lattice.}
\label{figmagic}
\end{figure}

The dipole potential of the 1D red-detuned optical lattice is constructed following Ref. \cite{Andrei-RMP-2011}. The maximum trapping depth at the antinode position of the lattice is determined by $U_0=-\alpha^{E1}_F(\omega){\cal E}^2$ with ${\cal E}=\sqrt{2I/(c \epsilon_0)}$, where $c$ and $\epsilon_0$ are speed of light and vacuum permittivity, respectively, and $I$ is average intensity of laser. The strength of the trapping potential is gauged by investigating the $\alpha^{E1}_F(\omega)$ values for the corresponding hyperfine levels of the considered states at $\lambda_m=1181$ nm. The $\alpha_F^{E1} (\omega)$ values of the levels associated with the pumping and repumping transitions are plotted for the wavelength range from 1140 nm to 1280 nm in Fig.
\ref{figmagic}. Using the $\alpha^{E1}_F(\omega)$ values, calculated in the same procedure as in Refs. \cite{cspol1,cspol2}, it gives $U_0$ for the $6S_{1/2}$ and $5D_{3/2,5/2}$ states, using a typical operational lattice intensity $I_{op}=20$ kW/cm$^2$, three times larger than the thermal kinetic energy $k_BT$ of the trapped atoms at the estimated temperature $T \sim 35 \ \mu$K and Boltzmann constant $k_B$. Due to presence of a large number of resonant
lines, it was strenuous to predict definite signs of $\alpha_F^{E1}(\omega)$ of the $7P_{3/2}$ state around $\lambda_m=1181$ nm. Also, the $U_0$
values for the $7S_{1/2}$ and $6P_{3/2}$ states turn out to be quite small. Nonetheless, atoms will decay from these states to the $6S_{1/2}$ and
$5D_{3/2,5/2}$ states within tens of nanoseconds before they could escape from the lattice.

Since it is imperative to operate the optical lattice at a very low sensitivity against variation in the trapping laser wavelength to minimize
fluctuation in the clock frequency, we estimate the rate of change in the clock frequency with respect to deviation in wavelength per unit power
of the trapping laser. This comes out to be of the order of 0.7 Hz/(GHz$\cdot$(kW/cm$^2$)) around $\lambda_m=1181$ nm from Fig. \ref{lasing}(c). By
assuming a typical experimental condition of maximum 100 kHz deviation in the trapping laser frequency \cite{Boydthesis}, the fractional clock
frequency shift is expected to be about $6.9\times10^{-18}$ for a typical value of $I_{op}=20$ kW/cm$^2$.

\begin{figure}[t]
\begin{center}
\includegraphics[width=8.0cm,height=4.5cm]{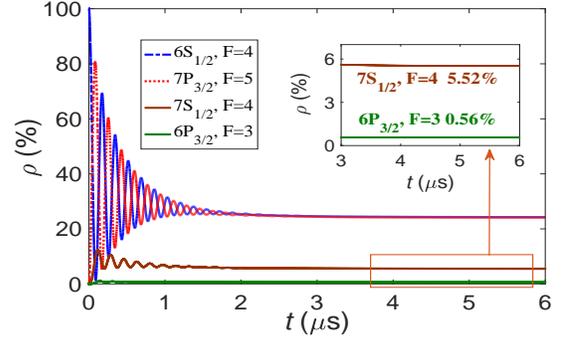}
\end{center}
\vspace{-6mm}
\caption{Steady-state atomic populations $\rho$ (in \%) in different levels over a range of time scale $t$ (in $\mu$s). Lattice, pumping, and repumping lasers are $\pi$-polarized, and the lattice laser is adjusted to be resonating with a mode of cavity.}
\label{popul}
\end{figure}

The steady-state atomic populations in the investigated four levels of Cs atom during the clock frequency measurement are obtained by considering
atom-pumping laser interactions in the Liouville or von Neumann equation, given by \cite{MarIan-quantum optics}
\begin{eqnarray}
\frac{{d \rho }}{{dt}} = \frac{1}{{i \hbar }} \left[ {H_{AP}, \rho } \right] - \frac{1}{2}\left\{ {\Gamma , \rho } \right\},\label{eq1}
\end{eqnarray}
where $H_{AP}=H^0_{AP}+H^I_{AP}$ is the total Hamiltonian comprising of the unperturbed Hamiltonian describing the upper ($| F_u, M_{F_u} \rangle$)
and the lower ($| F_l, M_{F_l} \rangle$) levels of the pumping (or repumping) transition as $H^0_{AP}= \sum_{i=u,l}  E_{F_i} | F_i, M_{F_i} \rangle
\langle F_i, M_{F_i}|$ with the eigenvalues $E_{F_i}$ in the absence of laser field and interaction Hamiltonian $H^I_{AP}= - \frac{i}{\hbar } \left
[ {\Omega_{F_lF_u}} | F_l, M_{F_l} \rangle\langle F_u, M_{F_u} |  +  {\Omega _{F_uF_l}} | F_u, M_{F_u} \rangle \langle F_l, M_{F_l} | \right]$
with the Rabi frequencies $\Omega_{F_lF_u}$= $\Omega_{F_uF_l}$ =$|\langle F_l\| D \|F_u \rangle| {\cal E}/\hbar$ for the field strength
${\cal E}$. We have used experimental values of energies and transition rates \cite{Cs.org,NIST} to predict the atomic populations, which are shown
in Fig. \ref{popul} for all the hyperfine levels of the above states and quoted explicitly in Table \ref{tab:magic} by taking the pumping and
repumping powers 500 mW/cm$^2$ and 2 mW/cm$^2$, respectively.

We also present the estimated lifetime $\tau_L$ and photon scattering rate $\Gamma_{sc}$ values in Table \ref{tab:magic}. The lifetime of an atomic state in the far-detuned optical lattice trap due to photon scattering is estimated by $\tau_L=U_0/ \Gamma_{sc}$, and the photon scattering rate $\Gamma_{sc}$ can be formulated as
\begin{equation}
\Gamma_{sc}=\frac{1}{6 (2J_n+1) }\sum_k \frac{ |\langle J_n || D || J_k \rangle|^2}{(\hbar \omega_L- \delta E_{nk} )^2}\Gamma_k{\cal E}^2,
\end{equation}
where $\Gamma_k$ is spontaneous decay rate from upper state $k$ to lower state $n$, and $\omega_L$ is lattice laser frequency.
As can be seen, the $\tau_L$ values are at least a few tens of seconds for different states, among which the $6P_{3/2}$ state has the shortest
lifetime. This is still long enough for continuous operation of the clock.

The output power of the superradiant laser in the atom-cavity coupled system is determined by solving the Born-Markov master equation
\cite{Meiser-PRL-2009,Kazakov-PRA-2017}, given by
\begin{eqnarray}
\frac{{d \rho }}{{dt}} = \frac{1}{{i\hbar }} \left[ {H_{AC},\rho } \right] + {\cal L} \left[ { \rho } \right]\label{eq2},
\end{eqnarray}
for upper $|F_u, M_{F_u} \rangle$ and lower $|F_l, M_{F_l} \rangle$ levels of the clock transition. Here, the total Hamiltonian of the
system is given by $H_{AC}=H^0_{AC}+H^I_{AC}$ with the unperturbed Hamiltonian $H^0_{AC}= \left [ \sum\limits_{j = 1}^{N_a} \frac{{\hbar {\omega_a}}}{2}
{\hat \sigma_j^z}  + \hbar {\omega_c}{\hat a_c^+ }\hat a_c \right ]$ and the interaction Hamiltonian $H^I_{AC}= \frac{{\hbar g }}{2}\sum\limits_{j = 1}^{N_a} ( {\hat a_c^+ }\hat \sigma^-_j + {\hat a_c^- }\hat \sigma^+_j)$. In these expressions, $\omega_a$ is the angular frequency of the clock transition, $\omega_c$ is the angular frequency of the cavity mode, $\hat \sigma_j^z=| F_u^j, M_{F_u}^j \rangle  \langle F_u^j, M_{F_u}^j| - | F_l^j, M_{F_l}^j \rangle  \langle F_l^j, M_{F_l}^j|$, $\hat \sigma_j^-=(\hat \sigma_j^+)^\dagger= | F_u^j, M_{F_u}^j \rangle  \langle F_l^j, M_{F_l}^j|$, atom-cavity coupling constant
$g= \mu \sqrt {\frac{8\pi\omega_a }{{\hbar {\varepsilon _0}V_c}}} =4.09 \times 10^6$ Hz for the actual E1 matrix element $\mu$ of the clock
transition and cavity mode volume $V_c$, and ${\cal L} \left[ { \rho } \right]$ is the net Liouvillian containing contributions from the cavity,
spontaneous decay, pumping laser and inhomogeneous lifetime of the upper state. The output power of the bad-cavity is estimated as
$P = \hbar {\omega _a}\kappa \ |{\rm tr}[\rho_{cw} a_c]|^2$ for the steady-state solution $\rho_{cw}$ of the above equation and cavity dissipation
rate $\kappa$. The estimated steady-state output power of the superradiant lasing is $P=24$ $\mu \rm{W}$ for a typical $N_a=10^7$ number of atoms achieved through population-inversion with the pumping and repumping laser intensities of 500 mW/cm$^2$ and 2 mW/cm$^2$, respectively.

The quantum restricted linewidth $\Delta v$, following the modified Schawlow-Townes formula for bad-cavity, is given by \cite{KuppenLinewidth}
\begin{equation}
\label{eq:linewidth}
\Delta v = \frac{{\hbar \omega _a {\kappa ^2}}}{{4\pi P}}{N_{sp}}\left( {1 + {{\left[ {\frac{{2({\omega _c} - {\omega _a})}}{{\Gamma_u  + \kappa }}} \right]}^2}} \right){\left( {\frac{\Gamma_u }{{\Gamma_u  + \kappa }}} \right)^2},
\end{equation}
where ${N_{sp}} = {\rho_u}/({\rho_u} - {\rho_l})$ with $\rho_u$ and $\rho_l$ being atomic population densities in upper and lower states, and
$\Gamma_u$ is the spontaneous decay rate of the clock transition. We have set the cavity dissipation rate as $\kappa = 100 \ \Gamma_u$. The
quantum-limit linewidth of the supperradiant lasing is anticipated to be narrowed down to 4.2 mHz for the $5.5\%$ and $0.6\%$ population densities
in the upper and lower levels of the clock transition, respectively.

\begin{table}[t]
\caption{\label{shift} Scalar and tensor components of E1 polarizabilities (in a.u.), and M1 polarizabilities (in a.u.) of the
clock levels. Estimated fractional shifts to the clock frequency from various systematics using the above quantities are listed
in the bottom part. \label{tab:alpha}}
{\setlength{\tabcolsep}{4pt}
\begin{tabular}{lrrrr}\hline\hline
Properties                  & \multicolumn{2}{c}{$|7S_{1/2} (F=4)\rangle$}  & \multicolumn{2}{c}{$|6P_{3/2} (F=3)\rangle$ }   \\ \hline \\
$\alpha^{E1(0)}_F$(0)       & \multicolumn{2}{r}{6238(15)}     & \multicolumn{2}{r}{1647(35)}                         \\
$\alpha^{E1(0)}_F$(455nm)   & \multicolumn{2}{r}{82(3)}        & \multicolumn{2}{r}{$-54(24)$}                          \\
$\alpha^{E1(0)}_F$(894nm)   & \multicolumn{2}{r}{$-267(11)$}   & \multicolumn{2}{r}{$-4129(111)$}                       \\
$\alpha^{E1(2)}_F$(455nm)   & \multicolumn{2}{r}{0}            & \multicolumn{2}{r}{$-3(1)$}                           \\
$\alpha^{E1(2)}_F$(894nm)   & \multicolumn{2}{r}{0}            & \multicolumn{2}{r}{1679(25)}                         \\
$\alpha^{M1}_F$             & \multicolumn{2}{r}{$-1.4(1)\times10^4$}      & \multicolumn{2}{r}{$-2.6(2)\times10^5$}     \\ \hline \\
                            \multicolumn{5}{c}{Fractional frequency shifts} \\
\multicolumn{3}{l}{BBR static at 300K}          &\multicolumn{2}{c}{1.94(2)$\times10^{-13}$}       \\
\multicolumn{3}{l}{455nm laser, $I$=500mW/cm$^2$}  &\multicolumn{2}{c}{$-1.6(7)\times10^{-14}$}    \\
\multicolumn{3}{l}{894nm laser, $I$=2mW/cm$^2$}   &\multicolumn{2}{c}{ $-1.8(1)\times10^{-15}$ }    \\
\multicolumn{3}{l}{$2^{nd}$-Zeeman, $\cal{B}$=$10^{-7}$ T} &\multicolumn{2}{c}{$-7.1(3)\times10^{-19}$} \\ \hline\hline
\end{tabular}}
\end{table}

Typical orders of magnitudes of the major systematics to the clock frequency due to black-body radiation (BBR) shifts, light shifts caused by the
pumping and repumping lasers, and the second-order Zeeman shifts, are determined by using the scalar ($\alpha^{\rm E1(0)}_{F}$) and tensor
($\alpha^{\rm E1(2)}_{F}$) E1 polarizabilities and magnetic dipole (M1) polarizabilities ($\alpha^{\rm M1}_{F}$) of the hyperfine levels that are
listed in Table \ref{tab:alpha}. Using the differential static scalar E1 polarizabilities, $\delta \alpha_F^{E1(0)}(0)$ between the $|7S_{1/2};F=4,
M_F=0\rangle$ and  $|6P_{3/2};F=3, M_F=0\rangle$ clock levels, the BBR shift at the room temperature (300 K) is determined by
\begin{equation}\label{eq:BBRE1}
\delta E^{\rm E1}_{\rm BBR}=-\frac{1}{2}(831.9{\rm~V/m})^2 \left [\frac{T(\rm K)}{300} \right]^4\delta {\alpha_{F}^{\rm E1(0)}}(0) .
\end{equation}
The fractional BBR shift in the clock frequency is estimated to be $1.94(2)\times10^{-13}$. This uncertainty can be suppressed by two more
orders by measuring the differential scalar E1 polarizability of the clock transition more precisely and conducting the experiment at a lower
temperature. The light shifts due to the pumping and repumping lasers lead to the fractional clock frequency shifts $-1.6(7)\times10^{-14}$ and
$-1.8(1)\times10^{-15}$, respectively. Since powers of these lasers can be controlled with an instability below $10^{-4}$ \cite{Rui2014Laser,
Tricot2018Power}, it is possible to curtail the above fractional uncertainties to the 10$^{-18}$ level. The choice of the $M_F=0\to M_{F}=0$ clock transition gives zero first-order Zeeman shift. The second-order Zeeman shift of a hyperfine level
is estimated by $\Delta E^{2^{nd}}_{\rm Zeem}=-\frac{1}{2} \alpha_{F}^{\rm M1} {\cal B}^2$. The fractional shift in the clock frequency due to this
effect is estimated from the difference of $\alpha^{\rm M1}_{F}$ values, which is determined as $\alpha_{F}^{\rm M1}= - \frac{2}{3(2F+1)}
\sum_{i} \frac{ |\langle F||O^{\rm M1}||F_i \rangle|^2}{E_{F}-E_{F_i}}$ for M1 operator $O^{\rm M1}$ and all possible $F_i$ levels of the clock
states with energies $E_{F_i}$. Uncertainties to these quantities come mainly from the energies. The $\alpha^{\rm M1}_{F}$ values of the
$|7S_{1/2}; F=4, M_F=0\rangle$ and $|6P_{3/2}; F=3, M_F=0\rangle$ levels are found to be $-1.4\times10^4$ a.u. and $-2.6\times10^5$ a.u.,
respectively. This gives a negligibly small fractional second-order Zeeman shift to the clock frequency. Other systematical effects are expected to be much lower than 10$^{-15}$ level.

In summary, we have proposed a continual superradiant lasing scheme to construct an active lattice clock based on optical lattices. A suitable
magic wavelength of the $|7S_{1/2}; F=4, M_F=0\rangle \rightarrow |6P_{3/2}; F=3, M_{F}=0 \rangle$ clock transition of Cs atom has been narrowed down to confine atoms strongly in the optical lattices with densely populated ground and excited states. By simulating realistic experimental
conditions, we predict to achieve fractional uncertainty to the clock frequency about 10$^{-15}$ level and linewidth of a few mHz. Its
bad-cavity operational mode can provide a robust short-term stability to this active clock. This in combination with another high-accuracy
passive optical clock can be used to cater better short-term and long-term stabilities for practical applications.

The work was supported by National Natural Science Foundation of China(NSFC)(91436210,11874064)and the Strategic Priority Research
Program of the Chinese Academy of Sciences (CAS), Grant No. XDB21030300, and the NKRD Program of China, Grant No. 2016YFA0302104. The work
of B.A. is supported by DST-SERB(India) Grant No. EMR/2016/001228. B.K.S. thanks IOP, Beijing for the overseas professor fellowship and
hospitality to carry out this work. D.P and J.B.C acknowledge valuable discussions with F. Fang.

\end{document}